\documentclass[runningheads]{llncs}

\usepackage[T1]{fontenc}

\usepackage{graphicx}

\usepackage{url}

\begin{document}
\title{Introductory Courses on Digital Twins: an Experience Report}

\author{
John S. Fitzgerald\inst{1}\orcidID{0000-0001-7041-1807} \and
Philip James\inst{1}\orcidID{0000-0001-9248-0280} \and 
Cl{\'a}udio Gomes\inst{2}\orcidID{0000-0003-2692-9742} \and 
Peter Gorm Larsen\inst{2}\orcidID{0000-0002-4589-1500}
}
\authorrunning{J. S. Fitzgerald, P. James, et al.}
\institute{Newcastle University, Newcastle upon Tyne NE1 7RU, UK 
\email{firstname.lastname@newcastle.ac.uk} \and
Dept. of Electrical and Computer Engineering, Aarhus University, DK
\email{\{claudio.gomes|pgl\}@ece.au.dk}
}
\maketitle              
\begin{abstract}
We describe and compare two new courses on model-based approaches to the engineering of Digital Twins. One course was delivered to doctoral students from a range of largely non-computational backgrounds, and the other to Masters students with computing experience. We describe the goals, content and delivery of the courses, and review experience gained to date. Key lessons focus on the importance of providing common baselines for participants coming from diverse technical backgrounds. 

\end{abstract}
\section{Introduction}
\label{sec:intro} 
Digital Twins (DTs) offer an attractive means to exploit models and data in a wide range of domains at levels from individual devices to systems-of-systems. Given the breadth of interest and potential in the DT concept, it is worth considering who will be engineering DT solutions, and what capabilities they require in order to take a systematic approach to DT Engineering. In the vision underpinning this paper, DT-based solutions will be developed and maintained by multidisciplinary teams, many of whose members are expert in application domains rather than underpinning computing and communications technology.  

Efforts such as Digital Twins as a Service (DTaaS) aim to raise the abstraction level of DT engineering by packaging elements of DTs as extensible building blocks~\cite{Talasila&25}. This has the goal of opening DT construction to a wider group of potential beneficiaries beyond those organisations with specific deep computing and communications skills.   However, even with such support, it is still necessary to equip DT engineers with the knowledge and skills to construct coherent DT solutions from these elements. We therefore consider the knowledge and skills that stakeholders from a wide range of backgrounds require, and hence how to offer meaningful training. 

This paper describes two recent efforts to develop and deliver initial education on DT engineering to two different cohorts of students in diverse settings. In Section~\ref{sec:background} we describe concepts basic to our underlying vision. In Section~\ref{sec:NCL} we describe a short course aimed at a multidisciplinary cohort of students training as researchers in a specific application domain (water infrastructure) that is facing increasing digitalisation. In Section~\ref{sec:AU}, we outline a technical course aimed at engineers creating DTs. We compare the experience of developing and delivering these courses in Section~\ref{sec:concl}.

\section{Background} 
\label{sec:background}

\subsection{DT Engineering and DT Engineers}
\label{sec:DTEngineering}

A DT is a system formed from one or more models that describe facets of a system of interest which we term the \emph{Physical Twin (PT)}\footnote{In many cases, in spite of the name, the PT may be cyber-physical-human in character.}. The PT and DT are linked by a \emph{communications infrastructure} that allows the DT to gather and analyse data from the PT. The models and gathered data form the foundations of services that add value by enabling capabilities that include visualisation, preventive maintenance, what-if analysis, adaptation and enhanced resilience. We use the term \emph{DT-enabled system} to refer to the system formed from the PT, communications infrastructure and DT together. 
\emph{DT Engineering} refers to the processes of analysing, specifying, designing, implementing, deploying, maintaining and disposing of DT-enabled systems. 

We regard DT Engineering as a systems engineering activity. The International Council on Systems Engineering (INCOSE) Systems Engineering Vision 2035 sees Digital Twins as part of global megatrends towards digital transformation changing products, processes and change strategies. However, it points out that most systems engineers have learned “on the job” and suggests that this limits the ability of such engineers to stay abreast of advances. It argues for education, training and lifelong learning that empowers  “more system engineers with strong multi- and transdisciplinary competencies” \cite{INCOSEvision2035}.

DT Engineering is inherently interdisciplinary in that DT-enabled systems may be developed by practitioners from a variety of cyber, physical and systems engineering backgrounds. This underpins the guidance and structuring built into the text recently edited by the authors~\cite{Fitzgerald&24}. However, we also anticipate that DT Engineering will be undertaken by teams containing varying mixtures of domain experts who will be familiar with a range of relevant system models alongside experts in the digital solution technologies of data and computation. The balance between these facets will often depend on the scale and capabilities of the stakeholder organisations. We therefore conclude that training on DT Engineering is needed for people from a widely varying disciplines and areas of domain expertise.  

We have begun to develop learning materials around the DT engineering concepts outlined in \cite{Fitzgerald&24}, including teaching content, course outlines and practical case study materials. We have taken the opportunity to deliver two courses based on the content developed to date, for two very different sets of students. At Newcastle University, the opportunity arose to work with a small cohort of early-career researchers-in-training who are domain experts but have limited computational expertise. At Aarhus University, almost the opposite situation arises, with a more technology-focussed course for engineers. In the following sections we describe each of these courses.  

\section{The Newcastle Course: Digitalisation in the Water Sector}\label{sec:NCL}

\subsection{Context}
\label{sec:NCLcontext}
The Newcastle course arose from the need to equip researchers, whether in industry or academia, with an awareness and basic knowledge of digitalisation. The domain in question was that of water infrastructure. 

Digitalisation is the process of transforming an activity by using digital technology, enabled by advances in sensing, communications, data science and computing capabilities. There is positivity around the opportunities afforded by digitalisation to overcome the greatest threats to the water supply~\cite{ECWater22}. However, there is also some caution that digital technologies provide a means to control and regulate the power relationships between the actors involved~\cite{Walter24}. 

The Newcastle course, offered through the Civil Engineering and Geosciences discipline within the School of Engineering, addresses the need for awareness of the role of digitalisation as a means of addressing challenges in water infrastructure, and specifically its resilience to a changing climate, patterns of supply and demand, and technologies. It was decided to do this using the DT as a vehicle for introducing a range of digitalisation concepts such as sensing and control, models, service-based system construction, etc. 

The specific context was the Centre for Doctoral Training (CDT) on Water Infrastructure Resilience~(WIRe\footnote{\url{https://cdtwire.com/}}). This collaboration of three UK universities (Cranfield, Sheffield and Newcastle) offers research training to cohorts of about 12 students annually. The programme lasts for four years, with an introductory year covering domain-specific learning at Masters level, and introductions to research principles, processes and skills. 

\subsection{Objectives and Structure}
\label{sec:NCLobj}
The course was developed as a single 5-credit (ECTS) module to be taken by all students in the cohort in the first semester of doctoral training as part of an ``induction'' suite of short courses. 

Given that these students would be embarking on doctoral (PhD) training in the water sector, and that some (but not all) of them would be undertaking projects with a digital component, we determined that the objective was to equip students to:  
\begin{itemize}
    \item Evaluate the potential of AI in the water sector
    \item Examine the fundamental concepts of Digital Twins from a Systems Engineering perspective
    \item Critique diverse models, data sources and analytic techniques 
    \item Debate the challenges of delivering a dependable digital twin in the water industry
    \item Appraise digital solutions
\end{itemize} 
Note here the focus on evaluation, examination, and critical appraisal rather than on solution construction. This was deliberate because our expectation was that these students may go on to specify and engage with digital solutions rather than develop them as an end in themselves. We judged that the capabilities outlined above would be of longer-term use to them than deep technical construction skills. 

All the outcomes would be assessed by group-based project work. Given the small scale of the module, the place of this module in the wider PhD programme, and the universities' current policies on assessment, this was preferred over a ``heavier'' process of examination-based individual assessment. 

\subsection{Content}
\label{sec:NCLcontent}

Given the module's objectives, the focus was on building an appreciation of the DT ``from the ground up'', with initial introductions to data and models, then building on this the communications element (linking models to the PT they claim to represent) and then the services built on these. Deeper practical training would be offered in data handling and AI. 

The theory content on DTs was based on the new text~\cite{Fitzgerald&24}. The practical data and AI content used examples drawn from a real-world water infrastructure research project. The Fair Water\footnote{\url{https://waterinnovation.challenges.org/winners/fair-water/}} project, tests and develops solutions to reducing carbon through energy and water efficiency, for example behaviour changes and product innovations. To date this has allowed water use data from sensors in 55 homes to (so far) be gathered over 10 months. This provided a relevant basis for the practical training elements. 

Assessed coursework was undertaken in groups. Students were asked to outline water-related DT solutions for one of the following scenarios: 
\begin{description}   
    \item[Group A: DT-enabled Homes] A housing developer, whose unique selling point is their engagement with technology and sustainability, is creating a pilot development in which each home's occupier has a DT of their house, and the developer has access to an estate-wide DT. Can a DT help the occupiers and developers to engage in sustainable water management practices? 
    \item[Group B: DT-enabled Campus] A university with a large urban estate (built between 1887 and 2024) needs to make strategic decisions about what areas to renovate and what areas to vacate. Can a DT help it take the most sustainable solutions from a water management perspective? 
    \item[Group C: DT-enabled City] A city in a delta is at risk of coastal, fluvial and pluvial flooding. DTs are promising but may assume access to high-tech (and high-finance!). Can we build an affordable, sustainable DT solution that enables informed decision-making in planning and flood management?
\end{description}
\noindent Each group's mission was to: 
\begin{enumerate}
    \item Research the potential of a DT solution, identify stakeholders and their needs. 
    \item Scope a DT solution by identifying a small number (max. 3) top-level services that should be offered to some or all stakeholders.
    \item Outline a DT solution following key elements of the systematic reporting framework for DT case studies identified by Gil et al.~\cite{Gil&24}, specifically:  
        \begin{itemize}
            \item System considerations: the DT scope, environment and stakeholders.
            \item Sources of data.
            \item Assets (Data and Models) to be gathered and maintained.
            \item The “constellation” of top-level services and intermediate enabler services required to deliver them. 
            \item Risks to the successful operation of the solution (stakeholder engagement; technology availability, dependability and security). 
            \item The life-cycle of the students' proposed solution: setting it up, maintenance and  decommissioning. 
        \end{itemize}
    \item Report by means of an online presentation followed by a question and answer session with two examiners and the class as a whole. Groups were required to ensure that all members of the group contributed to the presentation and to the discussion of their work. 
\end{enumerate}

\subsection{Delivery}
\label{NCLdelivery}
The module was first delivered in November/December 2024 to a group of 15 students mainly from the most recent cohort, plus several students from prior cohorts taking the module out of interest. Students had backgrounds including: chemistry, biology, civil and mechanical engineering and public policy. Students claimed only limited (but widely varying) levels of computing and data expertise.  

The module was delivered intensively over two weeks full-time. Students were present in person at Newcastle for Week 1, and worked remotely from their home institutions in Week 2. 

\begin{figure}[h!]
    \centering
    \includegraphics[width=0.9\linewidth]{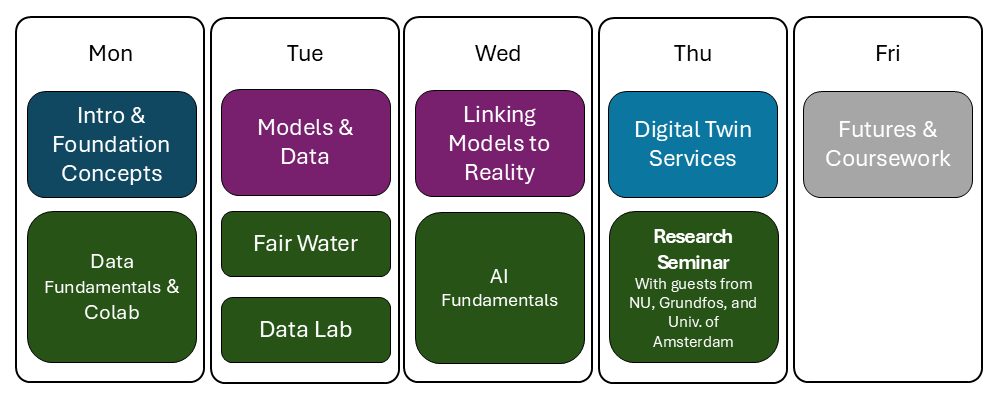}
    \caption{CEG8529 Week 1 Schedule}
    \label{fig:NUwk1schedule}
\end{figure}

The first week was structured as shown in Figure~\ref{fig:NUwk1schedule}. In the mornings, plenary classes introduced the material indicated in a traditional lecturing style, with discussion points and exercises integrated into these sessions. Afternoons generally included practice-based work. On Monday-Wednesday this took place at cluster computers and contained exercises aimed at increasing students' familiarity with other core aspects of digitalisation and AI. 

A notable feature of Week 1 was the inclusion of Thursday afternoon's seminar session which included presentations kindly offered by active practitioners and researchers in the field. Delivered remotely and in person, these covered  aspects of DT and broader digitalisation concepts in the water domain. One presentation (from Grundfos) addressed the significance and future of DTs and digitalisation in the sector from an industry perspective. One (from University of Amsterdam) was a full technical presentation of live research project activity in the area~\cite{Degeler24} and one (from Newcastle University) provided a broader research perspective. The talks were selected in order to provide industry validation of the significance of the material covered in the main lectures. The final day of Week 1 was devoted to a review of technical content, a forward look and an introduction to the assessed coursework. 

Week 2 was devoted to the group-based coursework. Three groups were formed of four or five students. Group membership was hand-picked in order to ensure, so far as we could, diversity among the individuals and their disciplinary backgrounds. All groups were cross-institutional so that the teams worked primarily online. Daily online drop-in sessions were arranged with the module leaders to allow discussion of ongoing work, refinement of the brief in each case, and to address any unanticipated events. 

\subsection{Experience}
\label{sec:NCLexperience}
Regarding setup and delivery, Week 1 was delivered in a flat-floor teaching space with desktop PCs at each desk, all forward facing. This was convenient for the afternoon practical classes but did not tend to promote easy discussion during lectures. On the Thursday afternoon and Friday, classes moved to another flat-floor space with moveable desks, which proved more suitable for the research seminar and group project setup sessions. 

The content of the lectures proved to have been reasonably well-judged. Each session had been constructed to ensure that there was some new material for students, whatever their background. There were generally good levels of student engagement and it was possible to gain confidence that core DT concepts were getting across. The afternoon practical sessions in Week 1 were considered broadly successful, but widely varying levels of digital maturity and software experience in the class meant that some students felt insufficiently challenged by the content. 

Regarding the assessed work, all three groups presented strong background research, conceptual overviews of the proposed DT-enabled systems,  identification of stakeholders, and acknowledgement of testing, security, and privacy concerns. Overall, the module leaders were encouraged to see a high level of ambition for data, digital technology, and AI in water infrastructure.   

The proposed DT-enabled systems all had clear overall purposes. However, the submissions were less strong at clearly defining an architecture for each DT showing the path from the gathered data and models to the enablers and top-level stakeholder-facing services, adding value to data on the way. Although models were discussed, the proposals lacked some detail on the characteristics (their type, purpose, fidelity, etc.) required to make the DTs function successfully.  The module leaders felt that there could have been more technical content describing the services needed to meet the DT’s purpose, as well as how these services were tiered and linked to provide a ‘logical’ system design to implement the conceptual one. We were impressed by the presentations, all of which were well-structured and clearly delivered. Students demonstrated confidence in addressing questions and handling challenges. All students who undertook the assessment completed the course successfully (some students studying for interest only did not require a grade). 

Student feedback was positive on the quality of content and presentation, and relevance to the water industry. Some students felt that the introductory coding content was too basic, some basic elements of successful cooperative group working were absent from one group (in the view of one member), and one student would have liked more general content on digitalisation. Several students recommended staying together in person for the group project element, and there is a case for greater emphasis being placed on successful team working.

\section{The Aarhus Course}
\label{sec:AU}

\subsection{Context: Engineering Digital Twins}
\label{sec:AUcontext}

At Aarhus University this course is offered for students at three different departments at the Masters level (Electrical and Computer Engineering (ECE), Mechanical and Production Engineering (MPE) and Civil and Architectural Engineering (CAE)). It is delivered as a 10-credit~(ECTS) course where the students are assessed with a combination of a group report and an oral examination. This course has a stronger focus on the underlying technology than the Newcastle course discussed in Section~\ref{sec:NCL}. For example, students are expected to demonstrate that they are able to engineer a DT with inspiration from the incubator DT example~\cite{Feng&21c}. 

\subsection{Objectives and Structure}
\label{sec:AUobj}
The objective was that students learn to engineer a DT, understanding the artefacts, models, calibration techniques, and communication methods required. They will also explore various DT services and gain hands-on experience through group projects. The learning outcomes are:

\begin{enumerate}
\item Identify and explain the necessary artefacts for a DT.
\item Compare different models of a PT and their pros and cons.
\item Evaluate calibration techniques for model fidelity.
\item Assess various methods for sensing and data communication from physical twins.
\item Discuss alternative DT services.
\item \label{AUoutcome} Engineer a complete DT system.
\end{enumerate}
\noindent where Objective~\ref{AUoutcome} is assessed by the group project and the others are assessed in an oral examination.

\subsection{Content}
\label{sec:AUcontent}

This course covers the essential components and techniques for developing and utilizing digital twins. It begins with an overview of DTs, their key elements, and models used for cyber-physical systems. Topics include state machines, Ordinary Differential Equations (ODEs), and methods for calibration. The course also addresses real-time data exchange, advanced services like software sensing, Kalman filtering, and deep learning, as well as 2D and 3D visualization tools. It further explores co-simulation, fault detection, system optimization, sensitivity analysis, uncertainty quantification, and the deployment of Digital Twins as a Service (DTaaS).

\subsection{Delivery}
\label{sec:AUdelivery}
\begin{figure}[h!]
    \centering
    \includegraphics[width=\linewidth]{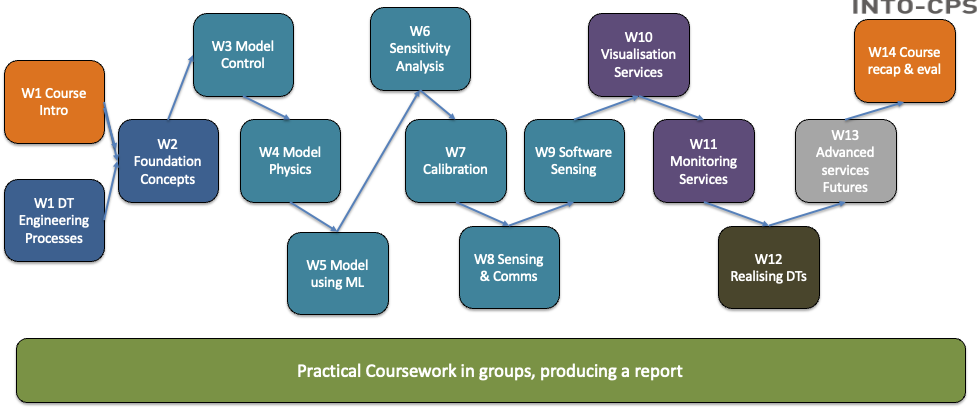}
    \caption{AU course 14 week schedule}
    \label{fig:AUwk1schedule}
\end{figure}

The first delivery of the course took place over the 14 weeks of the AU Spring semester 2025, with a class of 24 students. Delivery followed the structure shown in Figure~\ref{fig:AUwk1schedule}. Lectures (4 hours/week) were used to present underpinning theory while classes focussed on exercises (4 hours/week) gave students the opportunity to work in interdisciplinary groups to develop their own DT. Most students were from ECE, but a few students had backgrounds in mechanical and civil engineering. In order to address the varying needs of this mixed audience, Jupyter notebooks were prepared in advance for a range of topics that all the students would be expected to appreciate, including Git management, using Docker, using RabbitMQ and using the InfluxDB database. Each notebook contained optional exercises to help students engage with the content, enabling those without the necessary prerequisites to get up to speed quickly. The academic prerequisites for the course can be listed as:
\begin{enumerate}
\item Linear algebra.
\item Additional mathematics courses (e.g., discrete math and calculus).
\item General knowledge about engineering (e.g., systems engineering).
\item Basic programming skills (in e.g., C, C++, or Python).
\item Ideally experience in modelling either discrete or continuous systems.
\end{enumerate}

\noindent
A series of hand-ins are delivered by each group (with 5-6 students each):
\begin{enumerate}
\item Planning DT Requirements. 
\begin{itemize}
\item Write a plan for the DT of your case study. 
\item Understand the PT: what does it do? What are the physical processes occurring?
\item What is the model modelling?
\item What is an example parameter estimation task (calibration)?
\item What is an example fault that can be detected?
\item How to interact (set inputs/parameters, simulate, get outputs) with the model? How to run simulations? What are the allowed inputs?
\item What is an example machine learning task? How to collect the necessary data for it from the model or the real system? 
\item What is an example software sensing task? How can it be validated in a physical twin?
\end{itemize}
\item Modelling the PT. 
\item Present the model(s) created for the PT, and example applications of it.
\item Modelling the PT (Cont.) and First DT Service. 
\item Visualization and DT Monitoring Services. 
\item Final Report and Video Demonstration of DT Services. 
\end{enumerate}
Formative feedback is provided for all the intermediate hand-ins while the final report and video demonstrations received a summative assessment contributing the final course grade. Two of the groups worked with Universal Robots, one group worked with a hybrid test bench and one group worked a concrete batching plant. 

For the individual summative assessment, each student undertakes an oral examination in which they draw a topic at random and are asked to present on this for 15 minutes in dialogue with two examiners. The topics students can draw from are:
\begin{enumerate}
\item Describe and compare the processes needed for engineering DTs.
\item Compare alternative techniques for producing models inside DTs.
\item Explain how to calibrate models in a DT context.
\item Provide an overview of techniques for sensing and communicating between twins.
\item Provide an overview of visualisation techniques using in DTs.
\item Explain and compare monitoring capabilities inside DTs.
\item Reflect upon pros and cons with alternative ways of realising DTs.
\item Reflect upon the challenges of introducing autonomy in DTs.
\item Reflect upon the future research challenges for DTs.
\end{enumerate}

\subsection{Experience}
\label{sec:AUexperience}
The course was delivered for the first time in Spring 2025. The students were very engaged, and the feedback was overwhelmingly positive in terms of perceived importance to their studies, and a balanced workload. They successfully worked with the case studies and used the Jupyter notebooks to familiarize themselves with the various topics. However, the following challenges were identified:
\begin{itemize}
    \item The first practical lecture of the course should be used to ensure that all students fulfill the prerequisites. This can be achieved, for example, by providing mandatory exercises that require students to engage with the basic technologies. Our experience showed that students did not spend sufficient time outside of class to become familiar with the foundational tools and concepts.
    
    \item Starting with groups of five students was not ideal. One of the groups did not function well, and its members had to be reassigned, resulting in some groups having six students. Such large groups required a high level of coordination, which was challenging for members with less experience and led to disproportionate time spent on management. In future editions, we propose starting with groups of four students, allowing for easier redistribution in case one or two groups encounter difficulties.
    
    \item The different case studies had varying levels of maturity as starting points for the projects. For example, one case study included a physical twin emulator and a set of reusable models, enabling students to develop a comprehensive DT. Other case studies lacked a PT emulator and only provided models via an API, requiring students to implement the emulator themselves. This is because the case study providers had varying levels of capabilities with respect to digital to an engineering. This disparity resulted in some groups producing significantly more advanced DTs than others. In future editions, we recommend ensuring all case studies are provided at the same level of maturity.
    
    \item As the project progressed, the importance of having students with a software engineering background became increasingly evident due to the growing coding demands. We hypothesize that a good group composition might include two software engineering students for every student from another discipline.
\end{itemize}
\noindent At the time of writing, the final examinations have not taken place, and so the broad academic outcomes are yet to be determined.  

\section{Concluding Remarks}
\label{sec:concl}
We have outlined our first experiences in delivering two courses introducing DT Engineering concepts to diverse cohorts of students at Master's level. The Newcastle course aimed to equip engineering research students with an awareness of the potential of DTs and the ability to see through the ``hype'' on this subject. The Aarhus course aimed to provide deep technical skills encompassing DT construction and evaluation. On the basis of evidence to date, both courses were delivered successfully to highly engaged cohorts. 

At Newcastle, issues to address in the future include providing content that takes full account of differing levels of digital skills. We also felt that students were able to (but did not) engage more with DT solution architecture. We speculate that this was in part because they did not have opportunities to give semi-formal specifications of constituent models and services, and so in the future we will consider a rudimentary introduction to SysML as a course prerequisite delivered through preparatory material. Newcastle will also review the need to cover elements of successful group working that we had assumed students would already have before coming to the course.  

At Aarhus, future offerings will revisit the organisation and size of groups and providing a common level of case study maturity to ensure a consistent baseline for each group in the practical work. In order to better address the different starting points of students in terms of technical computing skills, we will place increased stress on the need to complete the Jupyter Notebooks ahead of time. 

Although based on a common core of materials (starting from \cite{Fitzgerald&24}), the two courses arguably focus on complementary facets of DT engineering processes, with the Newcastle course enabling domain experts to manage their expectations of the technology, and define potential DT-enabled systems while the Aarhus course focuses on the sound and effective creation of dependable DT-enabled systems. 

Finally, our experience suggests that there is positive benefit in offering a variety of courses for model-based DT Engineering, meeting the needs of different stakeholders working at diverse levels of abstraction. The authors have benefitted from developing a common repository of teaching and learning content, and would strongly encourage other teachers to join the repository (maintained courtesy of Aarhus University) to share their materials and experience in a range of settings. Those wishing to do so are invited to contact the authors directly. 

\noindent
\\
\small
\textbf{\ackname} We are grateful to many collaborators, including Justine Easten who organised the delivery of the Newcastle course and our pioneer students from Newcastle, Aarhus, Sheffield and Cranfield Universities. We thank the anonymous reviewers for their helpful comments on the first draft of this paper. We also gladly acknowledge the Grundfos Foundation's support to the AU Centre for Digital Twins. \textbf{\discintname} The authors have no competing interests to declare. 

\bibliographystyle{splncs04}
\bibliography{CEG8529}

\end{document}